%% file: paper.tex
\documentclass[fleqn,10pt]{wlscirep}

\usepackage[super]{nth}
\usepackage{rotating}
\usepackage{makecell}
\usepackage{pifont}
\usepackage{amsfonts}
\usepackage{amsmath}
\usepackage{float}
\usepackage{pgfplotstable}
\usepackage{booktabs,siunitx}

\pgfplotsset{compat=newest}
\usepackage{listings, textcomp}

\newcommand{\cmark}{\ding{51}}%
\newcommand{\code}[1]{\texttt{#1}}

\usepackage{etoolbox}
\makeatletter
\patchcmd{\@maketitle}
 {\noindent
{\parbox{\dimexpr\linewidth-2\fboxsep\relax}{\color{color1}\large\sffamily\textbf{ABSTRACT}}}}
 {\newpage\noindent
{\parbox{\dimexpr\linewidth-2\fboxsep\relax}{\color{color1}\large\sffamily\textbf{ABSTRACT}}}}
 {}{}
\makeatother

\title{SciPy 1.0---Fundamental Algorithms for Scientific Computing in Python}

\input{core-dev}


\begin{abstract}
SciPy is an open source scientific computing library for the Python programming language.
SciPy 1.0 was released in late 2017, about 16 years after the original
version 0.1 release. SciPy
has become a \emph{de facto} standard for leveraging scientific algorithms
in the Python programming language, with more than 600
unique code contributors, thousands of dependent packages,
over 100,000 dependent repositories, and millions of downloads per year.
This includes usage of SciPy
in almost half of all machine learning projects on GitHub, and usage by
high profile projects including LIGO gravitational wave analysis and creation
of the first-ever image of a black hole (M87).
The library includes functionality spanning clustering, Fourier transforms,
integration, interpolation, file I/O, linear algebra, image processing,
orthogonal distance regression, minimization algorithms, signal processing,
sparse matrix handling, computational geometry, and statistics. In this
work, we provide an overview of the capabilities and development practices of the
SciPy library and highlight some recent technical developments.
\end{abstract}
\begin{document}

\flushbottom
\maketitle
\thispagestyle{empty}

\section*{Introduction}




SciPy is a library of numerical routines for the Python programming
language that provides fundamental building blocks for modeling and
solving scientific problems. 
SciPy includes algorithms for optimization, integration, interpolation,
eigenvalue problems, algebraic equations, differential equations, and many other
classes of problems; it also provides
specialized data structures, such as sparse matrices
and $k$-dimensional trees. SciPy is built on top of 
NumPy\cite{oliphant2006guide, vanderwalt2011numpy},
which provides array data structures and related fast numerical routines, and
SciPy is itself the foundation upon which higher level scientific libraries,
including scikit-learn\cite{pedregosa2011scikit}
and scikit-image\cite{vanderwalt2014scikit}, are built. 
SciPy is relied upon by scientists, engineers, and others 
around the world. For example, published 
scripts\cite{alex_nitz_2018_1596771, LIGO-open}
used in the analysis of gravitational waves
\cite{PhysRevD.93.122003, abbott2017gw170817} 
import several subpackages of SciPy, and the M87 black
hole imaging project directly cites SciPy\cite{2019ApJ875L3E}.

Recently, SciPy released version 1.0, a milestone that traditionally
signals a library's API (Applications Programming Interface) being
mature enough to be be trusted in production pipelines. This version 
numbering convention, however, belies the history of a project that has
become the standard which others follow and has seen extensive
adoption in research and industry.

SciPy's arrival at this point is surprising and somewhat anomalous.
When started, the library had little funding, and was written mainly
by graduate students---many of them without a computer science education, and often without the
blessing of their advisors.  To even imagine that a small group of
``rogue'' student programmers could upend the already well-established
ecosystem of research software---backed by millions in funding and
many hundreds of highly qualified engineers\cite{mathworks-globe-97,
esri-revenue,bloom-wolfram}---was preposterous.

Yet the philosophical motivations behind a fully open tool stack, combined
with an excited, friendly community with a singular focus, have
proven auspicious in the long run.  They led not only to the library
described in this paper, but also to an entire ecosystem of related
packages \cite{scipy-ecosystem}, and a variety of social activities centered around
them\cite{social-python}. The packages in the SciPy ecosystem 
share high standards of implementation,
documentation, and testing, and a culture eager to learn and adopt
better practices---both for community management and software
development.


In the background section that follows, we capture a selective history
of some milestones and important events in the growth of SciPy.
Despite what we might highlight here, it is important to understand
that a project like SciPy is only possible because of the
contributions of very many contributors---too many to mention
individually, but each bringing an important piece to the puzzle.

\section*{Background}

Python is an interpreted, high-level, general-purpose computer programming
language, designed by Guido van Rossum in the late 1980s,
with a dynamic type system and an emphasis on readability and rapid prototyping.
The reference and most popular implementation of Python is 
CPython\cite{cpython-refman,cpython-source}, which is written
in the C and Python languages and assumed throughout this paper.
%
As a general purpose programming language, it had no special support for 
scientific data structures or algorithms, unlike many of the other established
computation platforms of the time. Yet, scientists soon discovered the
language's virtues, such as its ability to wrap C and Fortran
libraries and to then drive those libraries interactively.  Scientists
could thereby gain access to a wide variety of existing computational
libraries without concerning themselves with low-level programming
concepts such as memory management.

In 1995, Jim Hugunin, a graduate student from MIT, wrote the first
message in a new Python Matrix Special Interest Group (Matrix-SIG)
mailing list\cite{Hugunin-first}:
\begin{quote}
There seems to be a fair amount of interest in the Python community
concerning the addition of numeric operations to Python.  My own desire is
to have as large a library of matrix based functions available as possible
(linear algebra, eigenfunctions, signal processing, statistics, etc.).  In
order to ensure that all of these libraries interoperate, there needs to
be agreement on a basic matrix object that can be used to represent arrays
of numbers.
\end{quote}
Over the next several months, conversations on that mailing
list by, among others, Jim Fulton, Jim Hugunin, Paul Dubois, Konrad
Hinsen, and Guido van Rossum led to the creation of a package called Numeric with an array object
that supported a high number of dimensions.  Jim Hugunin explained the utility
of Python for numerical computation\cite{Hugunin-whitepaper}:
\begin{quote}
I've used almost all of the available numerical languages at one time
or another over the past 8 years. One thing I've noticed is that over
time, the designers of these languages are steadily adding more of the
features that one would expect to find in a general-purpose
programming language.
\end{quote}
This remains a distinguishing feature of Python for science, and one of the
reasons why it has been so successful in the realm of data science: instead of
adding general features to a language designed for numerical and scientific
computing, here scientific features are added to a general purpose language.  
This broadens the scope of problems that can be addressed easily, expands the
sources of data that are readily accessible, and increases the size of the
community that develops code for the platform.

The availability of a standard numerical array data structure in
Python led to a sudden, rapid growth in the number of scientific
packages available for solving common numeric problems.
A number of these packages were written by graduate students and
postdoctoral researchers to solve the very practical research problems
that they faced on a daily basis.  While they had access to and were
familiar with specialized (often commercial) systems, many found
it easier to implement the domain-specific functionality they needed
in a general purpose programming language.  And given Python's innate
ability to function as a systems language, controlling specialized or
custom-built hardware was also possible.


%


\subsection*{SciPy Begins}

By the late 1990s, discussions appeared on Matrix-SIG
expressing a desire for a complete scientific data analysis environment\cite{Travis-Keynote-2010}.
Travis Oliphant, a PhD student at the Mayo Clinic,
released a number of packages\cite{Travis-some-modules,Travis-enhance} 
that built on top of the Numeric array
package, and provided algorithms for signal processing, special
functions, sparse matrices, quadrature, optimization, Fast Fourier
Transforms, and more.  One of these packages, Multipack\cite{multipack}, was a set of
extension modules that wrapped Fortran and C libraries such as
ODEPACK\cite{hindmarsh1983odepack}, QUADPACK\cite{piessens1983quadpack}, and MINPACK\cite{osti_6997568}, to solve and minimize nonlinear
equations, integrate differential equations, and fit splines (the
latter implemented by Pearu Peterson).  Robert Kern, then an
undergraduate student (and currently a SciPy core developer), provided
compilation instructions under Windows.
Around the same time, Pearu Peterson, a PhD student from Estonia,
released F2PY\cite{peterson2009f2py}, a command line tool for binding Python and Fortran
codes, and wrote modules for linear algebra and interpolation.
Eric Jones, while a graduate student at Duke, wrote a number of
packages to support his dissertation, including a parallel job
scheduler and genetic optimizer.
%
%
Gary Strangman, a postdoctoral fellow at Harvard Medical School,
published several descriptive and inferential statistical 
routines\cite{Strangman-modules}.

\begin{figure}
\begin{verbatim}
SciPy is an open source package that builds on the strengths of Python and
Numeric providing a wide range of fast scientific and numeric functionality.
SciPy's current module set includes the following:

    Special Functions (Bessel, hanker, Airy, etc.)
    Signal/Image Processing
    2D Plotting capabilities
    Integration
    ODE solvers
    Optimization (simplex, BFGS, Netwon-CG, etc.)
    Genetic Algorithms
    Numeric -> C++ expression compiler
    Parallel programming tools
    Splines and Interpolation
    And other stuff.
\end{verbatim}
\caption{Excerpt from SciPy 0.1 release announcement (typos included).}\label{fig:announce-0.1}
\end{figure}

With a rich programming environment and a numerical array object in
place, the time was ripe for the development of a full scientific
software stack.
In 2001, Eric Jones and Travis Vaught founded Enthought Scientific
Computing Solutions (now Enthought, Inc.) in Austin, Texas.  In an
effort to simplify the tool stack, they created the SciPy project,
centered around the SciPy library which would subsume all the
above-mentioned packages.
The new project quickly gained momentum, with a website and code
repository appearing in
February\cite{archived-scipyorg}, and
a mailing list announced in
June \cite{new-scipy-list}. By
August, a first release was announced\cite{first-scipy-rel}, an excerpt of which is shown in
Fig.~\ref{fig:announce-0.1}.
In September, the first documentation was
published\cite{first-scipy-docs}.
The first SciPy
workshop\cite{first-scipy-workshop}
was held in September 2002 at Caltech---a single track, two day event with 50
participants, many of them developers of SciPy and surrounding libraries.



At this point, scientific Python started attracting more serious attention;
code that started out as side projects by graduate students had grown into
essential infrastructure at national laboratories and research institutes.
For example, Paul Dubois at Lawrence Livermore National Laboratory (LLNL) took over the
maintenance of Numeric and funded the writing of its
manual\cite{Numeric-manual}, and
the Space Telescope Science Institute (STScI), which was in charge of
Hubble Space Telescope science operations, decided to replace their
custom scripting language and analysis pipeline with Python\cite{STScI-slither}.

As SciPy, the algorithms library of the ecosystem, began attracting the attention
of large research organizations,
the next generation of graduate students and postdocs were already exploring
other aspects of the computational environment.
In 2000, Prabhu Ramachandran, a PhD student at the Indian Institute of
Technology, started work on a 3D visualization application, based on
Kitware's C++ Visualization Toolkit\cite{schroeder:2006:VTK}, called
Mayavi\cite{mayavi-intro}.
In 2001, Fernando Pérez, a graduate student at the University of
Colorado, Boulder, created the IPython interactive shell, which 
eventually blossomed into Project Jupyter\cite{Kluyver:2016aa}.
John Hunter, a postdoc
at the University
of Chicago, liked the plotting functionality
available in MATLAB\cite{matlab}, but had problems accessing their laboratory's
license, which was governed by a hardware dongle.  In response, he
wrote a plotting library from scratch, and Matplotlib 0.1 was released
April 2003\cite{matplotlib-rel}.



As STScI continued to use Python for an increasingly large portion
of the Hubble Space Telescope data analysis pipeline, they encountered
problems with the Python numerical array container.
Numeric, the original array package, was
suitable for small arrays, but not for the large images processed by
STScI.  With the Numeric maintainer's blessing, the decision was made
to write NumArray\cite{greenfield2003numarray}, a library that could handle data on a larger
scale.  Unfortunately, NumArray proved inefficient for small arrays,
presenting the community with a rather unfortunate choice.  In 2005,
Travis Oliphant combined the best elements of Numeric and NumArray,
thereby solving the dilemma and paving the way for Python to become
a very significant player in the Data Science movement. NumPy
1.0 was released in October 2006\cite{numpy-1.0-tag}, with about 30
authors recognized for major contributions in its release notes.


In a May/June 2007 special issue of IEEE Computing in Science and
Engineering, Paul Dubois wrote\cite{dubois2007guest}:
\begin{quote}
LLNL now has many Python-based efforts built from scratch or wrapped around
legacy codes [. . .] hundreds of thousands
of lines of C++, Python, and Fortran 95, all working together just as we hoped,
doing compute-intensive calculations on massively parallel computers.
\end{quote}
While it is now more commonly accepted, Dubois took the time to explain to
the 2007 reader that ``interpreted doesn't mean slow or only interactive.''
He also shared his own interest in Python for ``computational steering''
where ``Python serves as the input language to a scientific application, and the actual
computations are performed both in Python itself and in compiled extensions.''
Using Python for computational steering was one of the earliest uses of Python
in large scientific processing pipelines.
In addition to Dubois' overview there were also articles introducing
Python and SciPy \cite{oliphant2007python}, IPython \cite{perez2007ipython},
and  Matplotlib \cite{hunter2007matplotlib}.
There was also a diverse set of
research applications including
systems biology\cite{myers2007python},
astronomy\cite{greenfield2007reaching},
robotics\cite{krauss2007python},
nanophotonics\cite{bienstman2007python},
partial differential equations\cite{mardal2007using},
neuroimaging\cite{millman2007analysis},
geographic information systems\cite{shi2007python}, and
education\cite{backer2007computational, myers2007pythona}.

\subsection*{SciPy Matures}

By the middle to late 2000s, SciPy was starting to mature after a long phase of significant
growth and adoption.
The informal workshops grew into international conferences with many
hundreds of attendees. Special issues were organized and published in a
leading scientific journal\cite{dubois2007guest}, and the scope of the SciPy library
narrowed, while the breadth of the ecosystem grew
through a new type of auxiliary package: the scikit\cite{scikits-general}.
The tooling, development, documentation, and release processes became more professional.
SciPy was expanded carefully, with the patience affordable in open source 
projects and via best practices common in industry \cite{millman2014developing}.

Early versions of SciPy had minimal documentation, but this began to change
with the 2006 release of a \emph{Guide to NumPy}\cite{oliphant2006guide}. 
In 2007, Sphinx\cite{sphinx} made it possible to render hypertext and 
PDF-formatted documents automatically from plain text (docstrings) 
interspersed with Python code, and in 2008, \texttt{pydocweb}\cite{pydocweb}
enabled collaborative documentation development in a wiki-like fashion.
The SciPy Documentation Project \cite{vanderwalt2008scipy, harrington2008scipy} 
used these tools to complete documentation of SciPy's user-facing
functionality: offering t-shirts to contributors from around the world in 
exchange for high-quality text, it collected contributions from over 75 people 
to produce an 884 page manual\cite{harrington2009scipy}. Since then,
SciPy has remained committed to maintaining high-quality documentation as 
part of the normal development cycle.


In the early workshops, recurrent topics reflected the state of development, with
emphasis being placed on the underlying array package, plotting,
parallel processing, acceleration / wrapping, and user interfaces.  By
2004, a significant shift occurred towards application of SciPy to
scientific problems.  The event also started to draw in more keynote
speakers from outside the community, such as Guido van Rossum (creator
of Python, 2006), Ivan Krstić (One Laptop per Child, 2007), Alex
Martelli (Google, 2008), and Peter Norvig (Google Research, 2009).
The SciPy conference went from being a small gathering of
core developers to a multi-location event with increased funding, a
published proceedings, and scholarships for attending students.
By 2010, the US SciPy conference had multiple tracks, and satellite
conferences were being organized by volunteers elsewhere, such as EuroSciPy
(2008–) and SciPy India (2009–).
Special sessions and minisymposia dedicated to scientific Python began
appearing at many other events.
For example, a three-part minisymposium organized for CSE 2009 was
featured in SIAM News\cite{siamcse09}.

In 2007, Python had a strong enough presence in science and engineering
that the editors of IEEE Computing in Science and Engineering
(CiSE) solicited a special issue\cite{dubois2007guest}, edited by Paul
Dubois. However, Python was still sufficiently niche that the average reader
would need additional information in order to decide whether it would
``be useful in their own work.''
The follow-up CiSE March/April 2011 Python for Scientists and Engineers
special issue\cite{millman2011python} focused more on the core parts
of the scientific Python ecosystem\cite{perez2011python} including
NumPy\cite{vanderwalt2011numpy}, Cython\cite{behnel2011cython},
and Mayavi\cite{ramachandran2011mayavi}. Python became so pervasive that 
journals began publishing domain-specific special issues.
For example, in 2015 Frontiers in Neuroinformatics published a collection of 25 articles---covering
topics including modeling and simulation, data collection, electrophysiology, visualization,
as well as stimulus generation and presentation---called Python in
Neuroscience\cite{python-FIN}.

In 2012, Perry Greenfield, John Hunter, Jarrod Millman, Travis Oliphant,
and Fernando Pérez founded NumFOCUS\cite{numfocus},
 a 501(c)3 public charity with a mission
``to promote sustainable high-level programming languages, open code development,
and reproducible scientific research.''
While NumFOCUS is language agnostic, many of the early sponsored projects
came from the scientific Python stack.
Today it has 50 affiliated and sponsored projects including NumPy, SciPy, IPython, and
Matplotlib.
Among its other projects, NumFOCUS organizes a global network of 
community driven educational programs, meetups, and conferences
called PyData.

\subsection*{SciPy Today}
\begin{figure}[H]
\centering
\includegraphics[width=0.95\textwidth]{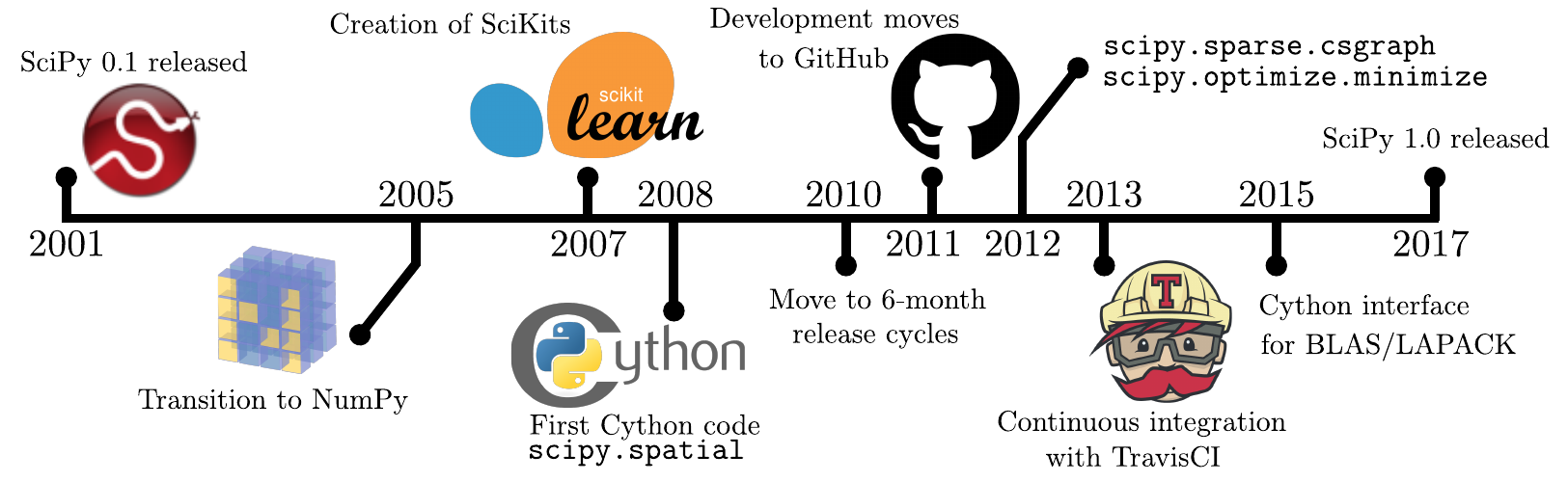}
\caption{Major milestones from SciPy's initial release in 2001 to
the release of SciPy 1.0 in 2017.}
\label{fig:timeline}
\end{figure}

At the time of writing, the SciPy library consists of nearly 
600,000 lines of code organized in 16 subpackages. 
Over 110,000 GitHub repositories and 6,500 packages depend
on SciPy\cite{dependents}. Some of the major
feature highlights from the three years preceding
SciPy 1.0 are discussed in the Key Technical Improvements section below,
and milestones in its history are highlighted in Figure~\ref{fig:timeline}.

\section*{Architecture and implementation choices}
\subsection*{Project scope}

SciPy provides fundamental algorithms for scientific computing. The
breadth of its scope was derived from the Guide to Available Mathematical
Software classification system (GAMS\cite{boisvert1991guide}). In areas
that move relatively slowly, e.g. linear algebra, SciPy aims to provide
complete coverage. In other areas it aims to provide fundamental building
blocks while interacting well with other packages specialized in that area.
For example, SciPy provides what one expects to find in a
statistics textbook (probability distributions, hypothesis tests, frequency
statistics, correlation functions, and more), while 
Statsmodels\cite{statsmodels2010} provides
more advanced statistical estimators and inference methods;
scikit-learn\cite{pedregosa2011scikit} covers machine learning; and
PyMC3\cite{10.7717/peerj-cs.55}, emcee\cite{2013PASP-emcee} and 
PyStan\cite{pystan-ref} cover Bayesian statistics and probabilistic modeling.
scikit-image\cite{vanderwalt2014scikit} provides image processing
capabilities beyond SciPy's \texttt{ndimage}, SymPy\cite{meurer2017sympy}
provides a Python interface for symbolic computation, and
while \texttt{sparse.csgraph} or \texttt{spatial} offers basic tools
for working with graphs and networks compared to a more specialized
Python library like NetworkX\cite{hagberg2008networkx}.

We use the following criteria to determine whether or not to include new
functionality in SciPy:
\begin{itemize}
    \item The algorithm is of relevance to multiple fields of science.
    \item The algorithm is demonstrably important.  For example, it is classic
    enough to be included in textbooks, or it is based on a peer-reviewed article
    which has a significant number of citations.
\end{itemize}

In terms of software systems and architecture, SciPy's scope matches NumPy's:
algorithms for in-memory computing on single machines, with support for a wide
range of data types and process architectures. Distributed computing and support
for graphics processing units (GPUs) are explicitly out of scope.

\subsection*{Package organization}
\input{subpackages}

\subsection*{Language choices}

According to analysis using the \texttt{linguist} library\cite{linguistref}, SciPy is approximately 50\% Python, 25\% Fortran, 20\% C, 3\% Cython, and 2\% C++, with a dash of \TeX, MATLAB, shell script, and Make. The distribution of secondary programming languages in SciPy is a compromise between a powerful, performance-enhancing language that interacts well with Python (i.e. Cython) and the usage of languages (and their libraries) that have proven reliable and performant over many decades.

Fortran, despite its age, is still a high-performance scientific programming language with 
continued contemporary usage\cite{Koelbel:1993:HPF:562354}. Thus, we wrap the following excellent, field-tested Fortran
libraries in order to provide Python convenience while benefiting from their performance: 
FFTPACK\cite{SWARZTRAUBER198445, SWARZTRAUBER198251}, 
ODEPACK\cite{hindmarsh1983odepack}, 
QUADPACK\cite{piessens1983quadpack}, 
FITPACK\cite{Dierckx:1993:CSF:151103}, 
ODRPACK\cite{ODRPACK_Boggs}, 
MINPACK\cite{osti_6997568}, 
ARPACK\cite{leh:sor:yan96}, 
ALGORITHM 644\cite{Amos:1986:APP:7921.214331}, and 
CDFLIB\cite{CDFLIB_site}. 

Rounding out the top three languages in SciPy is C, which is also extremely
well-established over several decades\cite{Kernighan:1988:CPL:576122} of
scientific computing. The C libraries that we wrap in SciPy include 
trlib\cite{doi:10.1080/10556788.2018.1449842}, 
SuperLU\cite{li05,superlu_ug99}, 
Qhull\cite{Barber:1996:QAC:235815.235821}, and 
Cephes\cite{cephes_netlib}. 

Cython has been described as a creole language that mixes the best parts of Python and
lower-level C / C++ paradigms\cite{behnel2011cython}. We often use Cython
as a glue between well-established, low-level scientific computing libraries written
in C / C++ and the Python interface offered by SciPy. We also use Cython to enable performance
enhancements in Python code, especially for cases where heavily used inner
loops benefit from a compiled code with static typing. 

For implementing new functionality, Python is the still the language of choice. If Python 
performance is an issue, then we prefer the use of Cython followed by C, C++, or Fortran (in that
order). The main motivation for this is maintainability: Cython has the 
highest abstraction level and most Python developers will understand it. C is 
also widely known, and easier for the current core development team to manage
than C++ and especially Fortran. 

The position that SciPy occupies near the
foundation of the scientific Python ecosystem is such that adoption of new
languages or major dependencies is generally unlikely---our choices are strongly
driven by long-term stability. GPU acceleration, new transpiling libraries, and
the latest JIT compilation approaches (i.e.,
Numba\cite{Lam:2015:NLP:2833157.2833162}) are very powerful, but currently fall
outside the remit of the main SciPy library. That said, we have recently
increased our efforts to support compatibility with some of these options, and
having our full test suite pass with the PyPy JIT
compiler\cite{Bolz:2009:TMP:1565824.1565827} is now a requirement in our
development workflow.

\subsection*{API and ABI evolution}

The application programming interface (API) for SciPy consists of approximately
1,500 functions and classes.  Our policy for evolving the API over time is that
new functionality can be added, while removing or changing existing
functionality can only be done if the benefits exceed the (often
significant) costs to users, \textit{and} only after giving clear deprecation
warnings to those users for at least one year. In general, we encourage
changes that improve clarity in the API of the library but strongly discourage
breaking backwards compatibility given our position near the base of the 
scientific Python computing stack.

In addition to the Python API, SciPy has C and Cython interfaces.
Therefore, we have to also consider the application binary
interface (ABI). This ABI has been stable for a long time, and we aim to
evolve it only in a backwards compatible way.

\section*{Key technical improvements}
\label{sec:technical_improvements}
Here we describe key technical improvements made in the last three years.

\subsection*{Data structures}

\subsubsection*{Sparse matrices}

\texttt{scipy.sparse} offers seven sparse matrix data structures,
also known as sparse formats. The most important ones are the row- 
and column-compressed formats (CSR and CSC, respectively). 
These offer fast major-axis indexing and fast matrix-vector multiplication,
and are used heavily throughout SciPy and dependent packages.

Over the last three years, our sparse matrix handling internals have been
rewritten and performance has been improved. Iterating over and slicing of CSC
and CSR matrices is now up to 35\% faster, 
and the speed of coordinate (COO) / diagonal (DIA) to CSR / CSC matrix format
conversions has increased. 
Importantly,
SuperLU\cite{superlu_ug99} was updated to version 5.2.1, enhancing the
low-level implementations leveraged by a subset of our \texttt{sparse}
offerings.

From a new features standpoint, \texttt{scipy.sparse} matrices and linear
operators now support the Python matrix multiplication (\texttt{@}) operator.
We've added \texttt{scipy.sparse.norm} and
\texttt{scipy.sparse.random} for computing sparse matrix norms and drawing
random variates from arbitrary distributions, respectively. Also, we've made a
concerted effort to bring the \texttt{scipy.sparse} API into line with the
equivalent NumPy API where possible.

\subsubsection*{\texttt{cKDTree}}

\input{ckdtree}

\subsection*{Unified bindings to compiled code}

\subsubsection*{LowLevelCallable}

As of SciPy version 0.19, it is possible for users to wrap low-level functions
in a \texttt{scipy.LowLevelCallable} object that reduces the overhead of
calling compiled C functions, such as those generated using \texttt{numba}
or Cython, directly from Python.
Supported low-level functions include \texttt{PyCapsule}
objects, \texttt{ctypes} function pointers, and \texttt{cffi} function pointers.
Furthermore, it is possible to generate a low-level callback function
automatically from a Cython module using \texttt{scipy.LowLevelCallable.from\_cython}.

\subsection*{Cython bindings for BLAS, LAPACK, and special}

SciPy has provided special functions and leveraged BLAS and
LAPACK\cite{LAPACK} routines for many years. SciPy now additionally
includes Cython\cite{behnel2011cython} wrappers for
many BLAS and LAPACK routines (added in 2015) and the special functions 
provided in the \texttt{scipy.{\allowbreak}special} subpackage (added in 2016).
These Cython wrappers are available in 
\texttt{scipy.{\allowbreak}linalg.{\allowbreak}cython\_blas},
\texttt{scipy.{\allowbreak}linalg.{\allowbreak}cython\_lapack}, and
\texttt{scipy.{\allowbreak}special.{\allowbreak}cython\_special} respectively.
When writing algorithms in Cython, it is typically more efficient to call
directly into the libraries SciPy wraps rather than indirectly, using SciPy's
Python APIs.  These low-level interfaces for Cython can also be used outside of
the SciPy codebase to gain access to the functions in the wrapped libraries
while avoiding the overhead of Python function calls.  This can give
performance gains of one or two orders of magnitude for many use cases.

Developers can also use the low-level Cython interfaces without linking against
the wrapped libraries\cite{blas-lapack-wrappers-scipy-2015}.  This lets other
extensions avoid the complexity of finding and using the correct libraries.
Avoiding this complexity is especially important when wrapping libraries
written in Fortran.  Not only can these low-level wrappers be used without a
Fortran compiler, they can also be used without having to handle all the
different Fortran compiler ABIs and name mangling schemes.

Most of these low-level Cython wrappers are generated automatically to help
with both correctness and ease of maintenance.  The wrappers for BLAS and
LAPACK are primarily generated using type information that is parsed from the
BLAS and LAPACK source files using F2PY\cite{peterson2009f2py}, though a small
number of routines use hand-written type signatures instead.  The input and
output types of each routine are saved in a data file that is read at build
time and used to generate the corresponding Cython wrapper files.  The wrappers
in \texttt{scipy.{\allowbreak}special.{\allowbreak}cython\_special} are also
generated from a data file containing type information for the wrapped
routines.

Since SciPy can be built with LAPACK 3.4.0 or later, Cython wrappers are only
provided for the routines that maintain a consistent interface across all
supported LAPACK versions.  The standard BLAS interface provided by the various
existing BLAS libraries is not currently changing, so changes are not generally
needed in the wrappers provided by SciPy.  Changes to the Cython wrappers for
the functions in \texttt{scipy.{\allowbreak}special} follow corresponding
changes to the interface of that subpackage.

\subsection*{Numerical optimization}

\input{scipy-optimize}

\subsection*{Statistical distributions}

The \texttt{scipy.stats} subpackage contains more than 100 probability
distributions: 96 continuous and 13 discrete univariate distributions
and 10 multivariate distributions. The implementation relies on a
consistent framework that provides methods to sample random variates,
to evaluate the cumulative distribution function (CDF) and the probability
density function (PDF) and to fit parameters for every distribution.
Generally, the methods rely on specific implementations for each
distribution such as a closed form expression of the CDF or a sampling
algorithm, if available. Otherwise, default methods are used
based on generic code, e.g., numerical integration of the PDF to
obtain the CDF.
Key recent distributions added to \texttt{scipy.stats} include the
histogram-based distribution in \texttt{scipy.stats.rv\_histogram}
and the multinomial distribution in \texttt{scipy.stats.multinomial}
(used, for example, in natural language processing, see
\cite{Griffiths5228}).

\subsection*{Polynomial interpolators}

\input{poly}

\subsection*{Test and benchmark suite}

\subsubsection*{Test suite}

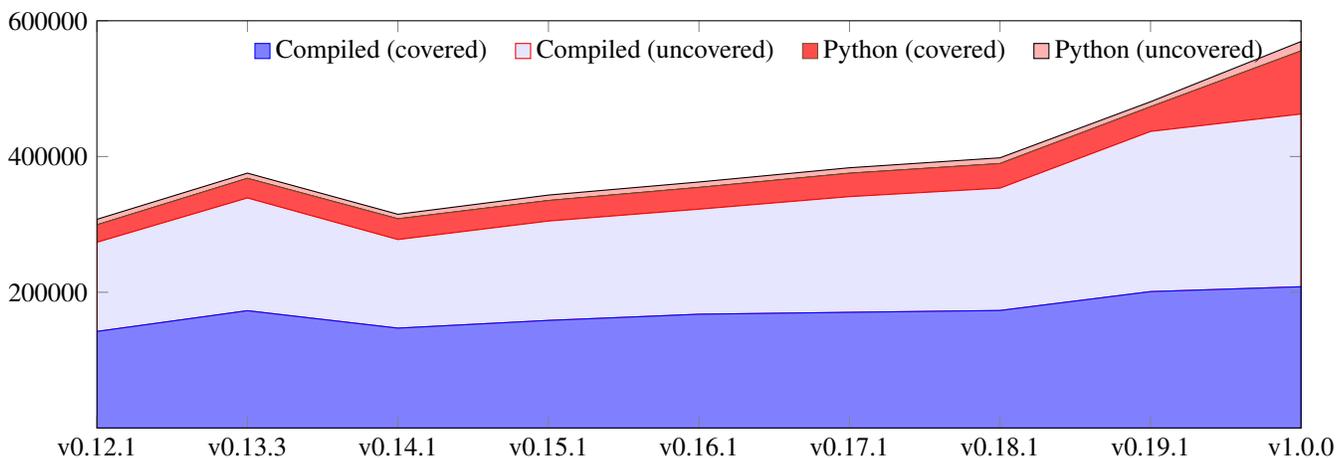
\begin{figure}[H]
\centering
\begin{tikzpicture}[]
\pgfplotstableread[column/ver/.style=string type]{
ver totpyline covpyline uncovpyline totcompline covcompline uncovcompline
v0.12.1 33844 25721 8123 273634  142290 131344
v0.13.3 36638 28944 7694 338926  172852 166074
v0.14.1 37301 30587 6714 277643  147151 130492
v0.15.1 38339 30288 8051 304898  158547 146351
v0.16.1 40094 32075 8019 322398  167647 154751
v0.17.1 42566 34478 8088 340903  170452 170451
v0.18.1 44711 36216 8495 353417  173174 180243
v0.19.1 43823 36373 7450 436900  200974 235926
v1.0.0 106878 92984 13894 462574 208158 254416
}\covtable
    \begin{axis}[
    width=\textwidth,
    height=7cm,
    enlargelimits=false,
    ymin=0,
    ymax=6e5,
    stack plots=y,%
    area style,
    xtick=data,
    ytick={2e5,4e5,6e5},
    xticklabels from table={\covtable}{ver},
    scaled y ticks=false,
    y tick label style={/pgf/number format/.cd, fixed, 1000 sep={}},
    legend entries={Compiled (covered),Compiled (uncovered),Python (covered),Python (uncovered)},
    legend style={draw=none,fill=none, cells={align=left},/tikz/every even column/.append style={column sep=3mm}},
    legend columns=-1,
    legend image code/.code={\fill[#1] (0cm,-0.1cm) rectangle (0.2cm,0.1cm);},
    ]
  \addplot+[fill=blue!50] table[x expr=\coordindex, y=covcompline] {\covtable} \closedcycle;
  \addplot+[fill=blue!10] table[x expr=\coordindex, y=uncovcompline] {\covtable} \closedcycle;
  \addplot+[fill=red!70] table[x expr=\coordindex, y=covpyline] {\covtable} \closedcycle;
  \addplot+[fill=red!30] table[x expr=\coordindex, y=uncovpyline] {\covtable} \closedcycle;
\end{axis}
\end{tikzpicture}
\caption{
Python (red) and compiled (blue) code volume in SciPy over time.
Deep-shaded area represents lines of code covered by units tests;
light-shaded area represents lines not covered. With the exception
of the removal of $\approx 61,000$ lines of compiled code for SciPy
v0.14, the volume of both compiled (C, C++, and Fortran) and Python
code has increased between releases, as has the number of lines
covered by unit tests. In SciPy v1.0.0, tests covered approximately
45\% of $462,574$ lines of compiled code (many of which
are automatically generated from Cython source), and nearly 87\% of
106,878 lines of Python code.\\
Coverage analysis is performed automatically in our continuous integration
suite using the \texttt{pytest-cov} and \texttt{gcov} libraries for 
Python and compiled code, respectively.
The reconstitution of historical build and test environments for this
figure was performed using a Docker-based approach\cite{scipy-cov}.
}
\label{fig:coverage}
\end{figure}

The SciPy test suite is orchestrated by a continuous integration matrix that
includes POSIX and Windows (32/64-bit) platforms managed by Travis CI and
AppVeyor, respectively. Our tests cover Python versions 2.7, 3.4, 3.5, 3.6, and
include code linting with \texttt{pyflakes} and \texttt{pycodestyle}. There
are more than $13,000$ unit tests in the test suite, which is written for usage
with the \texttt{pytest}\cite{pytest} framework.
Test coverage at the SciPy 1.0 release point (Figure~\ref{fig:coverage}) was at
87\% for Python code according to \texttt{pytest-cov}\cite{pytest-cov} and
45\% for compiled (C, C++, and Fortran) code according to \texttt{gcov}\cite{gcov}.
Documentation for the code is automatically built and published by
the CircleCI service to facilitate evaluation of documentation changes /
integrity.  Our full test suite also passes with PyPy3\cite{Bolz:2009:TMP:1565824.1565827}, a just-in-time compiled
version of the Python language.

\subsubsection*{Benchmark suite}

\begin{figure}[H]
\centering
\includegraphics[width=\textwidth]{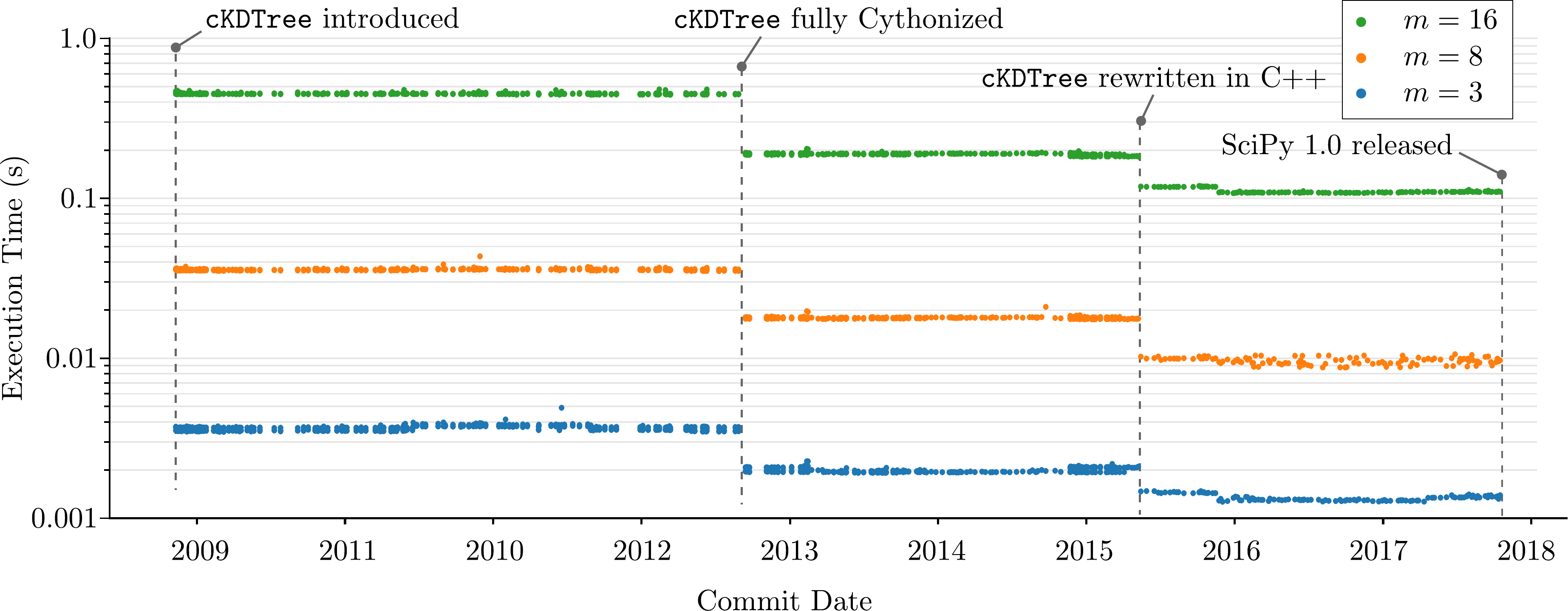}
\caption{Results of the \texttt{scipy.spatial.cKDTree.query} benchmark from the introduction of \texttt{cKDTree} to the release of SciPy 1.0. The benchmark generates a $k$-d tree from 10,000 uniformly distributed points in an $m$-dimensional unit hypercube, then finds the nearest (Euclidean) neighbor in the tree for each of 1,000 query points. Each marker in the figure indicates the execution time of the benchmark for a commit in the master branch of SciPy. Substantial performance improvements were realized when \texttt{cKDTree} was fully Cythonized and again when it was rewritten in C++. The tree was generated without application of toroidal topology (\texttt{boxsize=None}), and tests were performed by Airspeed Velocity 0.4 using Python 2.7, NumPy 1.8.2, and Cython versions 0.27.3, 0.21.1, and 0.18 (for improved backward compatibility).}
\label{fig:asvbench}
\end{figure}

In addition to ensuring that unit tests are passing, it is important to confirm that the
the performance of the SciPy codebase improves over time. Since February 2015, the
performance of SciPy has been monitored with Airspeed Velocity (\texttt{asv} \cite{asvref}).
SciPy's \texttt{run.py} script conveniently wraps \texttt{asv} features such that benchmark 
results over time can be generated with a single console command. For example,
Figure~\ref{fig:asvbench} illustrates the improvement of\texttt{scipy.spatial.cKDTree.query}
over roughly nine years of project history.

\section*{Project organization and community}

\subsection*{Governance}

SciPy adopted an official Governance Document on August 3, 2017\cite{SciPyProjectGovernance}. A Steering Council, currently composed of 18 members, oversees daily development of the project by contributing code and reviewing contributions from the community. Council Members have commit rights to the project repository, but they are expected to merge changes only when there are no substantive community objections. The Chair of the Steering Council, Ralf Gommers, is responsible for initiating biannual technical reviews of project direction and summarizing any private Council activities to the broader community. The project's Benevolent Dictator for Life (BDFL), Pauli Virtanen, has overruling authority on any matter, but the BDFL is expected to act in good faith and only exercise this authority when the Steering Council cannot reach agreement.

SciPy's official Code of Conduct was approved on October 24, 2017. In summary, there are five Specific Guidelines:
\emph{be open} to everyone participating in our community;
\emph{be empathetic and patient} in resolving conflicts;
\emph{be collaborative}, as we depend on each other to build the library;
\emph{be inquisitive}, as early identification of issues can prevent serious consequences; and
\emph{be careful with wording}.
The Code of Conduct specifies how breaches can be reported to a Code of Conduct Committee, and outlines procedures for the Committee's response. Our Diversity Statement ``welcomes and encourages participation by everyone''.

\subsection*{Maintainers and contributors}

The SciPy project has approximately 100 unique contributors
for every 6-monthly release cycle. Anyone with the interest and
skills can become a contributor; the SciPy Developer
Guide\cite{scipy-dev-guide} provides guidance on how to do that.
In addition, the project currently has 15 active maintainers: people who review
the contributions of others and do everything else needed to ensure that the
software and the project move forward. Maintainers are critical to the health
of the project\cite{eghbal2016}; their skills and efforts largely determine how
fast the project progresses, and they enable input from the much
larger group of contributors. Anyone can become a maintainer, too, as they
are selected on a rolling basis from contributors with a significant history of
high-quality contributions.

\subsection*{Downstream projects}

The scientific Python ecosystem includes many examples
of domain-specific software libraries building on top
of SciPy features and then returning to the base SciPy library
to suggest and even implement improvements. 
For example, there are common contributors to the SciPy and 
Astropy core libraries\cite{astropy-2018}, and what works 
well for one of the codebases, infrastructures, or communities
is often transferred in some form to the other. At the codebase
level, the \texttt{binned\_statistic} functionality
is one such cross-project contribution: it was initially
developed in an Astropy-affiliated package
and then placed in SciPy afterwards. 
In this perspective, SciPy serves as a catalyst for cross-fertilisation 
throughout the Python scientific computing community.

\section*{Discussion}


SciPy has a strong developer community and a massive user base. GitHub traffic
metrics report roughly 20,000 unique visitors to the source website between May
14 and May 27, 2018, with 721 unique copies (``clones") of the codebase over
that roughly two-week period. The developer community at that time consisted of 610 unique
contributors of source code, with $>19,000$ commits accepted into the codebase
(GitHub page data).

From the user side, there were 13,096,468 downloads of SciPy from the Python
Packaging Index (PyPI)\cite{pypinfo} and 5,776,017 via the default channel of the
\texttt{conda}\cite{condainfo} package manager during the year 2017. These numbers establish a lower
bound on the total number of downloads by users given that
PyPI and \texttt{conda} are only two of several popular methods for installing SciPy.  The SciPy
website\cite{SciPylib}, which has been the default citation in the absence of a
peer-reviewed paper, has been cited $>3,000$ times\cite{googlescholar}. Some of the most prominent
usages of or demonstrations of credibility for SciPy include the LIGO / Virgo
scientific collaboration that lead to the observation of gravitational waves
\cite{PhysRevLett.116.061102}, the fact that SciPy is shipped directly with
macOS and in the Intel Distribution for Python\cite{intel-python}, and that SciPy is used
by 47\% of all machine learning projects on GitHub\cite{octoverse-scipy}.

Nevertheless, SciPy continually strives to improve.
The SciPy Roadmap\cite{SciPy_roadmap_1,SciPy_roadmap_dev}, summarized in Table~\ref{table:roadmap}, is a
continuously-updated document
maintained by the community that describes some of the major directions for
improvement for the project, as well as specific limitations and matters that
require assistance moving forward.
In addition to the items on the roadmap, 
we are still working to increase the number of SciPy usage tutorials beyond
our current 15 section offering\cite{SciPy_tutorials}. 
Also, the low-level Cython code in our library (which interacts with C-level code and
exposes it for Python usage) could use some measure of modernization, including
migration to typed memoryviews to handle NumPy arrays.

%

\begin{table}[H]
\caption{Summary of SciPy Roadmap items following 1.0 release}
\begin{tabular}{ c|c }
    SciPy subpackage & Summary of Change \\
    \hline
    \texttt{optimize} & a few more high quality global optimizers\\
    \texttt{fftpack} & reduce overlap with NumPy equivalent \\
    \texttt{linalg} & reduce overlap with NumPy equivalent \\
    \texttt{interpolate} & new spline fitting and arithmetic routines \\
    \texttt{interpolate} & new transparent tensor-product splines\\
    \texttt{interpolate} & new Non-Uniform Rational B-Splines\\
    \texttt{interpolate} & mesh refinement \& coarsening of B-splines and tensor products\\
    \texttt{signal} & migrate spline functionality to \texttt{interpolate}\\
    \texttt{signal} & Second Order Sections update to match capabilities in other routines\\
    \texttt{linalg} & support a more recent version of LAPACK\\
    \texttt{ndimage} & clarify usage of the ``data point'' coordinate
    model, and add additional wrapping modes\\
    \texttt{sparse} & incorporate sparse arrays from Sparse package\cite{abbasi2018sparse} \\
    \texttt{sparse.linalg} & add PROPACK wrappers for faster SVD\\
    \texttt{spatial} & add support for (quaternion) rotation matrices\\
    \texttt{special} & precision improvements for hypergeometric, parabolic cylinder, and spheroidal wave functions\\
\end{tabular}\label{table:roadmap}
\end{table}

A problem faced by many open source projects is attracting and retaining developers. While it is normal for some individuals to contribute to a project for a while and then move on, too much turnover can result in the loss of institutional memory, leading to mistakes of the past being repeated, APIs of new code becoming inconsistent with the old code, and a drifting project scope.
We are fortunate that the SciPy project continues to attract enthusiastic and competent new developers while maintaining the involvement of a small but dedicated ``old guard''. There are contributors who were present in the early years of the project who still contribute to discussions of bug reports and reviews of new code contributions. Our BDFL has been with the project for more than 10 years and is still actively contributing code, and the head of our Steering Council, who also acts as a general manager, is approaching his eleventh anniversary. An additional half dozen or so active developers have been contributing steadily for five or more years. The combination of a committed old guard and a host of new contributors ensures that SciPy will continue to grow while maintaining a high level of quality.

\section*{Data Availability}
All SciPy library source code and most data generated for the current study are available in the SciPy GitHub repository, \url{https://github.com/scipy}. Some supporting code and data have also been stored in other public repositories cited by this manuscript.

\bibliography{references}

\section*{Acknowledgements}

We thank everyone who has contributed to SciPy 1.0, from those who have posted one comment about an issue through those who have made several small patches and beyond!

\section*{Author Contributions Statement}

P.V. is SciPy's Benevolent Dictator for Life, R.G. is the Chair of the SciPy Steering Council, and together they have led the development of SciPy for over 10 years. T.E.O., E.J., and P.P. created SciPy. T.J.R. and M.H. composed the manuscript with input from others. Other named authors are SciPy Core Developers. All authors have contributed significant code, documentation, and/or expertise to the SciPy project. All authors reviewed the manuscript.


\section*{Competing Interests}

The authors declare no competing interests.

\section*{Consortium}
\subsection*{SciPy 1.0 Contributors}
\input{consortium}

\end{document}

%% file: core-dev.tex
\author[1]{Pauli Virtanen}
\author[2,*]{Ralf Gommers}
\author[3,4,5,6,2]{Travis E. Oliphant}
\author[7,8,*]{Matt Haberland}
\author[9,*]{Tyler Reddy}
\author[10]{David Cournapeau}
\author[11]{Evgeni Burovski}
\author[12,13]{Pearu Peterson}
\author[10]{Warren Weckesser}
\author[14]{Jonathan Bright}
\author[15]{St\'efan J. van der Walt}
\author[16]{Matthew Brett}
\author[10]{Joshua Wilson}
\author[15,17]{K. Jarrod Millman}
\author[18]{Nikolay Mayorov}
\author[19]{Andrew R. J. Nelson}
\author[5]{Eric Jones}
\author[5]{Robert Kern}
\author[20]{Eric Larson}
\author[21]{CJ Carey}
\author[10]{\.{I}lhan Polat}
\author[22]{Yu Feng}
\author[23]{Eric W. Moore}
\author[24]{Jake VanderPlas}
\author[10]{Denis Laxalde}
\author[10]{Josef Perktold}
\author[25]{Robert Cimrman}
\author[26]{Ian Henriksen}
\author[10]{E. A. Quintero}
\author[10]{Charles R Harris}
\author[27]{Anne M. Archibald}
\author[28]{Ant\^{o}nio H. Ribeiro}
\author[29]{Fabian Pedregosa}
\author[30]{Paul van Mulbregt}
\author[ ]{SciPy 1.0 Contributors}
\affil[1]{University of Jyv\"askyl\"a, FI-40014 University of Jyv\"askyl\"a, Finland}
\affil[2]{Quansight LLC, Austin, TX, USA}
\affil[3]{Ultrasound Imaging, Mayo Clinic, Rochester, MN 55902, USA}
\affil[4]{Electrical Engineering, Brigham Young University, Provo, UT 84604, USA}
\affil[5]{Enthought, Inc., Austin, TX 78701, USA}
\affil[6]{Anaconda Inc, Austin, TX 78701, USA}
\affil[7]{BioResource and Agricultural Engineering Department, California Polytechnic State University, San Luis Obispo, CA 93407, USA}
\affil[8]{Department of Mathematics, University of California, Los Angeles, CA 90025, USA}
\affil[9]{Los Alamos National Laboratory, Los Alamos, NM 87544}
\affil[10]{Independent Researcher}
\affil[11]{Higher School of Economics, Myasnitskaya 20, Moscow 101000 Russia}
\affil[12]{Independent Researcher, Saue, Estonia}
\affil[13]{Department of Mechanics and Applied Mathematics, Institute of Cybernetics at Tallinn Technical University, Akadeemia Rd 21 Tallinn, Estonia}
\affil[14]{Independent Researcher, NY, USA}
\affil[15]{Berkeley Institute for Data Science, University of California, Berkeley, CA 94720, USA}
\affil[16]{School of Psychology, University of Birmingham, Edgbaston, Birmigham B15 2TT, UK}
\affil[17]{Division of Biostatistics, University of California, Berkeley, CA 94720, USA}
\affil[18]{WayRay LLC, Skolkovo Innovation Center, Moscow, 143026, Russia}
\affil[19]{Australian Nuclear Science and Technology Organisation, Locked Bag 2001, Kirrawee DC, NSW 2232, Australia}
\affil[20]{Institute for Learning and Brain Sciences, University of Washington, Seattle, WA 98195, USA}
\affil[21]{College of Information and Computing Sciences, University of Massachusetts Amherst, Amherst, MA 01002, USA}
\affil[22]{Berkeley Center for Cosmological Physics, University of California, Berkeley, CA 94720, USA}
\affil[23]{Bruker Biospin Corp., 15 Fortune Drive, Billerica, MA 01821}
\affil[24]{University of Washington, Seattle, WA 98195, USA}
\affil[25]{New Technologies - Research Centre, University of West Bohemia, Univerzitní 8, 306 14, Plzeň, Czech Republic}
\affil[26]{Oden Institute for Computational Engineering and Sciences, The University of Texas at Austin, Austin, Texas 78712, USA}
\affil[27]{Anton Pannekoek Institute, P.O. Box 94249, 1090 GE Amsterdam, The Netherlands}
\affil[28]{Graduate Program in Electrical Engineering, Universidade Federal de Minas Gerais, 6627 Av. Antonio Carlos 31270-901 Belo Horizonte, Brazil}
\affil[29]{Google LLC, Montreal, QC H3B 2Y5, Canada}
\affil[30]{Google LLC, 355 Main St., Cambridge, MA 02142, USA}
\affil[*]{scipy.articles@gmail.com}

%% file: subpackages.tex
The SciPy library is organized as a collection of subpackages.  The 16
subpackages include mathematical building blocks (e.g. linear algebra, Fourier
transforms, special functions), data structures (e.g. sparse matrices, $k$-D trees),
algorithms (e.g. numerical optimization and integration, clustering, interpolation,
graph algorithms, computational geometry), and higher-level data analysis
functionality (e.g. signal and image processing, statistical methods).

Here we summarize the scope and capabilities of each subpackage.

\begin{description}[leftmargin=!, labelwidth=\widthof{\bfseries \texttt{interpolate}}]
\itemsep0em
\item[\texttt{cluster}]
    The \texttt{cluster} subpackage contains \texttt{cluster.vq}, which provides vector quantization and $k$-means algorithms, and \texttt{cluster.hierarchy}, which provides functions for hierarchical and
    agglomerative clustering.
\item[\texttt{constants}]
    Physical and mathematical constants, including the CODATA recommended
    values of the fundamental physical constants\cite{CODATA2014}.
\item[\texttt{fftpack}]
    Fast Fourier Transform routines.  In addition to the FFT itself, the subpackage
    includes functions for the discrete sine and cosine transforms and for
    pseudo-differential operators.
\item[\texttt{integrate}]
    The \texttt{integrate} subpackage provides tools for the numerical
    computation of single and multiple definite integrals and for the
    solution of ordinary differential equations, including initial value
    problems and two-point boundary value problems.
\item[\texttt{interpolate}]
    The \texttt{interpolate} subpackage contains spline functions and
    classes, one-dimensional and multi-dimensional (univariate and
    multivariate) interpolation classes, Lagrange and Taylor polynomial
    interpolators, and wrappers for FITPACK\cite{Dierckx:1993:CSF:151103} and DFITPACK functions.
\item[\texttt{io}]
    A collection of functions and classes for reading and writing MATLAB\cite{matlab}, IDL,
    Matrix Market\cite{boisvert1997matrix}, Fortran, NetCDF\cite{rew1990netcdf},
		Harwell-Boeing\cite{harwellboeing}, WAV and ARFF data files.
\item[\texttt{linalg}]
    Linear algebra functions, including
    elementary functions of a matrix, such as the trace, determinant, norm and
    condition number;
    basic solver for $Ax = b$;
    specialized solvers for Toeplitz matrices, circulant matrices, triangular
    matrices and other structured matrices; least squares solver and
    pseudo-inverse calculations; eigenvalue and eigenvector calculations
    (basic and generalized); matrix decompositions, including Cholesky, Schur,
    Hessenberg, $LU$, $LDL^{\intercal}$, $QR$, $QZ$, singular value, and polar;
    and functions to create specialized matrices, such as diagonal, Toeplitz,
    Hankel, companion, Hilbert, and more.
\item[\texttt{ndimage}]
    This subpackage contains various functions for multi-dimensional image
    processing, including convolution and assorted linear and nonlinear
    filters (Gaussian filter, median filter, Sobel filter, etc.);
    interpolation; region labeling and processing; and mathematical morphology
    functions.
\item[\texttt{misc}]
    A collection of functions that did not fit into the other subpackages.
    While this subpackage still exists in SciPy 1.0, effort is underway
    to deprecate or relocate the contents of this subpackage and remove it.
\item[\texttt{odr}]
    Orthogonal distance regression, including Python wrappers for the Fortran
    library ODRPACK\cite{ODRPACK_Boggs}.
\item[\texttt{optimize}]
		This subpackage includes simplex and interior-point linear programming
		solvers, implementations of many nonlinear minimization algorithms,
		a routine for least-squares curve fitting, and 
		a collection of general nonlinear solvers for root-finding.
\item[\texttt{signal}]
    The \texttt{signal} subpackage focuses on signal processing and
    basic linear systems theory.  Functionality includes
    convolution and correlation, splines, filtering and filter design,
    continuous and discrete time linear systems, waveform generation,
    window functions, wavelet computations, peak finding, and spectral
    analysis.  
\item[\texttt{sparse}]
    This subpackage includes implementations of several representations of
    sparse matrices.\newline 
		\texttt{scipy.sparse.linalg} provides a collection of linear algebra routines that work with
		sparse matrices, including linear equation solvers, eigenvalue decomposition, singular value
		decomposition and LU factorization.
    \texttt{scipy.sparse.csgraph} provides a collections of graph algorithms
    for which the graph is represented using a sparse matrix.  Algorithms
    include connected components, shortest path, minimum spanning tree
    and more.
\item[\texttt{spatial}]
    This subpackage provides spatial data structures and algorithms,
    including the $k$-d tree, Delaunay triangulation, convex hulls and Voronoi
    diagrams. \texttt{scipy.spatial.distance} provides
    a large collection of distance functions, along with functions for
    computing the distance between all pairs of vectors in a given collection
    of points or between all pairs from two collections of points.
\item[\texttt{special}]
    The name comes from the class of functions traditionally known as ``special
    functions'', but over time, the subpackage has grown to include functions
    beyond the classical special functions.  A more appropriate characterization
    of this subpackage is simply ``useful functions''.
    It includes a large collection of the classical special functions
    such as Airy, Bessel, etc.; orthogonal families of polynomials;
    the Gamma function, and functions related to it;
    functions for computing the PDF, CDF and quantile function for several
    probability distributions;
    information theory functions;
    combinatorial functions \texttt{comb} and \texttt{factorial};
    and more.
\item[\texttt{stats}]
    The \texttt{stats} subpackage provides a large collection of continuous
    and discrete probability distributions, each with methods to compute
    the PDF or PMF, CDF, moments and other statistics, generation of random
    variates, and more;
    statistical tests, including tests on equality of means/medians/variance
    (such as the t-test) and tests whether a sample is drawn from a
    certain distribution (such as the Kolmogorov-Smirnov test);
    measures of correlation, including Pearson's $r$, Kendall's $\tau$, and
    Spearman's $\rho$ coefficients;
    descriptive statistics including trimmed values;
    kernel density estimation;
    and transformations of data such as the Box-Cox power transformation.
\end{description}

%% file: ckdtree.tex
The \texttt{scipy.spatial.ckdtree} module, which implements a space-partitioning data structure that
organizes points in $k$-dimensional space, was rewritten in C++ with templated classes. 
Support was added for periodic boundary conditions, which are often used 
in simulations of physical processes. 

In 2013, the time complexity of the $k$-nearest neighbor search from
\texttt{cKDTree.query} was approximately loglinear \cite{knn-jake},
consistent with its formal description \cite{kdtree-search-algo}.
Since then, we've enhanced \texttt{cKDTree.query} by reimplementing it in
C++, removing memory leaks, and allowing release of the global interpreter lock (GIL) so that
multiple threads may be used\cite{gh-4374}. This generally improved
performance on any given problem (Figure~\ref{fig:asvbench})
while preserving the asymptotic complexity.

In 2015, SciPy added the \texttt{sparse\_distance\_matrix} routine for generating 
approximate sparse distance matrices between \texttt{KDTree} objects by ignoring 
all distances that exceed a user-provided value. This routine is not 
limited to the conventional L2 (Euclidean) norm but supports any Minkowski 
$p$-norm between 1 and infinity. By default, the returned data structure is a
Dictionary Of Keys (DOK) based sparse matrix, which is very efficient for matrix 
construction. This hashing approach to sparse matrix assembly can be 7 times 
faster than constructing with CSR format
\cite{10.1007/978-3-540-75755-9_107}, and the C++ level sparse matrix construction 
releases the Python GIL for increased performance. Once the matrix is constructed, 
distance value retrieval has an amortized constant time complexity 
\cite{Cormen:2001:IA:580470}, and the DOK structure can be efficiently converted 
to a CSR, CSC, or COO matrix to allow for 
speedy arithmetic operations.

In 2015 the \texttt{cKDTree} dual tree counting algorithm\cite{Moore2000ar}
was enhanced to support weights\cite{ckdtree-weights}, which are
essential in many scientific applications, e.g. computing correlation
functions of galaxies\cite{0004-637X-750-1-38}.

%% file: scipy-optimize.tex
\newcommand{\RR}{\ensuremath{\mathbb{R}}}
The \texttt{scipy.optimize} subpackage provides functions for the numerical
solution of several classes of root finding and optimization problems.
Here we highlight recent additions through SciPy 1.0.


\subsubsection*{Linear Optimization}

A new interior-point optimizer for continuous linear programming problems, \texttt{linprog} with \texttt{method='interior-point'}, was released with SciPy 1.0. Implementing the core algorithm of the commercial solver MOSEK \cite{andersen2000mosek}, it solves all of the 90+ NETLIB LP benchmark problems \cite{netlib} tested. Unlike some interior point methods, this homogeneous self-dual formulation provides certificates of infeasibility or unboundedness as appropriate. 

A presolve routine \cite{andersen1995presolving} solves trivial problems and otherwise performs problem simplifications, such as bound tightening and removal of fixed variables, and one of several routines for eliminating redundant equality constraints is automatically chosen to reduce the chance of numerical difficulties caused by singular matrices. Although the main solver implementation is pure Python, end-to-end sparse matrix support and heavy use of SciPy's compiled linear system solvers---often for the same system with multiple right hand sides due to the predictor-corrector approach---provide speed sufficient for problems with tens of thousands of variables and constraints.

Compared to the previously implemented simplex method, the new interior-point method is faster for all but the smallest problems, and is suitable for solving medium- and large-sized problems on which the existing simplex implementation fails. However, the interior point method typically returns a solution near the center of an optimal face, yet basic solutions are often preferred for sensitivity analysis and for use in mixed integer programming algorithms. This motivates the need for a crossover routine or a new implementation of the simplex method for sparse problems in a future release, either of which would require an improved sparse linear system solver with efficient support for rank-one updates.

\subsubsection*{Nonlinear Optimization}
\paragraph{Local Minimization}
The \texttt{minimize} function provides a unified interface for finding local minima of nonlinear optimization problems. Four new methods for unconstrained optimization were added to \texttt{minimize} in recent versions of SciPy: \texttt{dogleg}, \texttt{trust-ncg}, \texttt{trust-exact}, and \texttt{trust-krylov}. All are trust-region methods that build a local model of the objective function based on first and second derivative information, approximate the best point within a local ``trust region'', and iterate until a local minimum of the original objective function is reached, but each has unique characteristics that make it appropriate for certain types of problems. For instance, \texttt{trust-exact} achieves fast convergence by solving the trust-region subproblem almost exactly, but it requires the second derivative Hessian matrix to be stored and factored every iteration, which may preclude the solution of large problems ($\geq 1000$ variables). On the other hand, \texttt{trust-ncg} and \texttt{trust-krylov} are well-suited to large-scale optimization problems because they do not need to store and factor the Hessian explicitly, instead using second derivative information in a faster, approximate way. A detailed comparison of the characteristics of all \texttt{minimize} methods is presented in Table~\ref{tab:minimize-si}; it illustrates the level of completeness that SciPy aims for when covering a numerical method or topic.

\paragraph{Global Minimization}
As \texttt{minimize} may return any local minimum, some problems require the use of a global optimization routine. The new \texttt{scipy.optimize.differential\textunderscore evolution} function \cite{Wormington1999,Storn1997} is a stochastic global optimizer that works by evolving a population of candidate solutions. In each iteration, trial candidates are generated by combination of candidates from the existing population. If the trial candidates represent an improvement, then the population is updated. Most recently, the SciPy benchmark suite gained a comprehensive set of 196 global optimization problems for tracking the performance of existing solvers over time and for evaluating whether the performance of new solvers merits their inclusion in the package.

\setlength{\tabcolsep}{3pt}
\begin{table}[H]
  \centering
  \caption{Optimization methods from \texttt{minimize}, which solves problems of the form $\min_x f(x)$, where $x \in \mathbb{R}^n$ and $f: \mathbb{R}^n \rightarrow \mathbb{R}$ .  The field \textit{version added} specifies the algorithm's first appearance in SciPy. Algorithms with \textit{version added} ``0.6*'' were added in version 0.6 or before.
    The field \textit{wrapper} indicates whether the implementation available in SciPy wraps a function written in a compiled language
    (e.g. C or FORTRAN). The fields \textit{\nth{1}} and \textit{\nth{2} derivatives}
    indicates whether first or second order derivatives are required. When \textit{\nth{2} derivatives} is flagged
    with $\sim$ the algorithm does not require second-order derivatives from
    the user; it computes an approximation internally and uses it to accelerate method convergence.
    \textit{Iterative Hessian factorization} denotes algorithms that factorize the Hessian in an iterative way,
    which does not require explicit matrix factorization or storage of the Hessian.
    \textit{Local convergence} gives a lower bound on the rate of convergence of the iterations sequence once the
    iterate is sufficiently close to the solution: linear (L), superlinear (S) and quadratic (Q). Convergence rates denoted S$^*$ indicate that the algorithm
    has a superlinear rate for the parameters used in SciPy, but can  achieve a quadratic convergence rate with other parameter choices.
    \textit{Global convergence} is marked for the algorithms with guarantees of convergence to a stationary
    point (i.e. a point $x^*$ for which $\nabla f(x^*) = 0$); this is \emph{not} a guarantee of convergence to a global minimum. The table also indicates which algorithms
    can deal with constraints on the variables. We distinguish among \textit{bound constraints} (i.e. $x^l \le x \le x^u$),
    \textit{equality constraints} (i.e. $c_{\text{eq}}(x) = 0$) and \textit{inequality constraints} (i.e. $c_{\text{ineq}}(x) \ge 0$).}
  \begin{tabular}{cccccccccccccc}
      & \rotatebox{80}{\texttt{Nelder-Mead}} & \rotatebox{80}{\texttt{Powell}} & \rotatebox{80}{\texttt{COBYLA}} & \rotatebox{80}{\texttt{CG}} & \rotatebox{80}{\texttt{BFGS}}&  \rotatebox{80}{\texttt{L-BFGS-B}} & \rotatebox{80}{\texttt{SLSQP}} & \rotatebox{80}{\texttt{TNC}} & \rotatebox{80}{\texttt{Newton-CG}} & \rotatebox{80}{\texttt{dogleg}} & \rotatebox{80}{\texttt{trust-ncg}} & \rotatebox{80}{\texttt{trust-exact}} & \rotatebox{80}{\texttt{trust-krylov}} \\
    \hline
    Version added &  0.6* &  0.6* &  0.6* &  0.6* &  0.6* &  0.6* &  0.9 &  0.6* &  0.6* & 0.13 & 0.13 & 0.19 & 1.0 \\
    \hline
    Wrapper & & & \cmark & & & \cmark & \cmark & \cmark & &  & & & \cmark \\
    \hline
    \nth{1} derivatives &  & & & \cmark  & \cmark & \cmark & \cmark & \cmark & \cmark & \cmark & \cmark & \cmark & \cmark \\
    \hline
    \nth{2} derivatives &  &  &  &  & $\sim$ & $\sim$ & $\sim$ & \cmark & \cmark & \cmark & \cmark & \cmark & \cmark \\
    \hline
    \makecell{Iterative Hessian \\
    factorization} & & & &  & & & & \cmark & \cmark &  & \cmark &  & \cmark \\
    \hline
    Local convergence& & & & L & S &  L & S & S$^*$ & S$^*$ & Q & S$^*$ & Q & S$^*$  \\
    \hline
    Global convergence & & &  &   & \cmark & \cmark & \cmark & \cmark & \cmark & \cmark & \cmark & \cmark & \cmark  \\
    \hline
    \makecell{Line-search (LS) or\\ trust-region (TR)} & Neither  & LS &  TR & LS & LS & LS & LS & LS & LS & TR & TR & TR & TR \\
    \hline
    Bound constraints &&&\cmark&&&&\cmark&\cmark&\cmark&&&& \\
    \hline
    Equality constraints &&&&&&&\cmark&&&&& \\
    \hline
    Inequality constraint &&&\cmark&&&&\cmark&&&&& \\
    \hline
    References & \cite{nelder_simplex_1965, wright_direct_1996} & \cite{powell_efficient_1964} &
      \cite{powell_direct_1994, powell_direct_1998, powell_view_2007} &
      \cite{polak_note_1969, nocedal_numerical_2006} & \cite{nocedal_numerical_2006} & \cite{byrd_limited_1995, zhu_algorithm_1997} &
      \cite{schittkowski_nonlinear_1982, schittkowski_nonlinear_1982-1, schittkowski_convergence_1983, kraft_software_1988} &
      \cite{nash_newton-type_1984} & \cite{nocedal_numerical_2006}  & 
      \cite{powell_new_1970, nocedal_numerical_2006} &  \cite{steihaug_conjugate_1983, nocedal_numerical_2006} &
      \cite{conn_trust_2000, more_computing_1983} & \cite{gould_solving_1999, doi:10.1080/10556788.2018.1449842} \\
    \hline
  \end{tabular}
  \label{tab:minimize-si}
\end{table}

%% file: poly.tex
Historically, SciPy relied heavily on the venerable FITPACK
Fortran library by P.~Dierckx \cite{Dierckx:1993:CSF:151103, FITPACK} for
univariate interpolation and approximation of data, but the original monolithic
design and API for interaction between SciPy and FITPACK was limiting for both
users and developers.

Implementing a new, modular design of polynomial interpolators was spread over
several releases. The goals of this effort were to have a set of basic objects
representing piecewise polynomials, to implement a collection of algorithms
for constructing various interpolators, and to provide users with building
blocks for constructing additional interpolators.

At the lowest level of the new design are classes which represent univariate
piecewise polynomials: \code{PPoly} (SciPy 0.13)\cite{scipy-gh2885},
\code{BPoly} (SciPy 0.13) and \code{BSpline} (SciPy 0.19)\cite{scipy-gh3174},
which allow
efficient vectorized evaluations, differentiation, integration and root-finding.
\code{PPoly} represents piecewise polynomials in the power basis in terms of
breakpoints and coefficients at each interval. \code{BPoly} is similar, and
represents piecewise polynomials in the Bernstein basis (which is suitable
for e.g., constructing Bezier curves). \code{BSpline} represents spline
curves, i.e., linear combinations of B-spline basis elements.\cite{deBoor1978} 

In the next layer, these polynomial classes are used for constructing several
common ways of interpolating data: \code{CubicSpline} (SciPy 0.18)
\cite{scipy-gh5653} constructs a twice 
differentiable piecewise cubic function, \code{Akima1DInterpolator} 
and \code{PCHIPInterpolator} implement two classic prescriptions for
constructing a $C^1$ continuous monotone shape-preserving interpolator.
\cite{FritschCarlson1980, Akima1970}

%% file: consortium.tex
{\bfseries
Aditya Vijaykumar$^{31,32}$, 
Alessandro Pietro Bardelli$^{10}$, 
Alex Rothberg$^{10}$, 
Andreas Hilboll$^{33}$, 
Andreas Kloeckner$^{34}$, 
Anthony Scopatz$^{2}$, 
Antony Lee$^{35}$, 
Ariel Rokem$^{36}$, 
C. Nathan Woods$^{9}$, 
Chad Fulton$^{37}$, 
Charles Masson$^{38}$, 
Christian Brueffer$^{39}$, 
Christian H\"aggstr\"om$^{40}$, 
Clark Fitzgerald$^{41}$, 
David A. Nicholson$^{42}$, 
David R Hagen$^{43}$, 
Dmitrii V. Pasechnik$^{44}$, 
Emanuele Olivetti$^{45}$, 
Eric Martin$^{10}$, 
Eric Wieser$^{46}$, 
Fabrice Silva$^{47}$, 
Felix Lenders$^{48,49,50}$, 
Florian Wilhelm$^{51}$, 
G. Young$^{10}$, 
Gavin A. Price$^{52}$, 
Gert-Ludwig Ingold$^{53}$, 
Gregory E. Allen$^{54}$, 
Gregory R. Lee$^{55,56}$, 
Herv\'e Audren$^{57}$, 
Irvin Probst$^{58}$, 
J\"org P. Dietrich$^{59,60}$, 
Jacob Silterra$^{61}$, 
James T Webber$^{62}$, 
Janko Slavi\v{c}$^{63}$, 
Joel Nothman$^{64}$, 
Johannes Buchner$^{65,66}$, 
Johannes Kulick$^{67}$, 
Johannes L. Sch\"{o}nberger$^{10}$, 
Jos\'e Vin\'icius de Miranda Cardoso$^{68}$, 
Joscha Reimer$^{69}$, 
Joseph Harrington$^{70}$, 
Juan Luis Cano Rodr\'iguez$^{71}$, 
Juan Nunez-Iglesias$^{72}$, 
Justin Kuczynski$^{73}$, 
Kevin Tritz$^{74}$, 
Martin Thoma$^{75}$, 
Matthias K\"ummerer$^{76}$, 
Maximilian Bolingbroke$^{77}$, 
Michael Tartre$^{78}$, 
Mikhail Pak$^{79}$, 
Nathaniel J. Smith$^{10}$, 
Nikolai Nowaczyk$^{10}$, 
Nikolay Shebanov$^{80}$, 
Oleksandr Pavlyk$^{81}$, 
Per A. Brodtkorb$^{82}$, 
Perry Lee$^{10}$, 
Robert T. McGibbon$^{83}$, 
Roman Feldbauer$^{84}$, 
Sam Lewis$^{85}$, 
Sam Tygier$^{86}$, 
Scott Sievert$^{87}$, 
Sebastiano Vigna$^{88}$, 
Stefan Peterson$^{10}$, 
Surhud More$^{89,90}$, 
Tadeusz Pudlik$^{91}$, 
Takuya Oshima$^{92}$, 
Thomas J. Pingel$^{93}$, 
Thomas P. Robitaille$^{94}$, 
Thomas Spura$^{10}$, 
Thouis R. Jones$^{95}$, 
Tim Cera$^{10}$, 
Tim Leslie$^{10}$, 
Tiziano Zito$^{96}$, 
Tom Krauss$^{97}$, 
Utkarsh Upadhyay$^{98}$, 
Yaroslav O. Halchenko$^{99}$, 
Yoshiki V\'azquez-Baeza$^{100}$
}
\newline
\hfill \break
$^{31}$International Centre for Theoretical Sciences, Tata Institute of Fundamental Research, Bengaluru 560089, India, 
$^{32}$Department of Physics, Birla Institute of Technology and Science, Pilani 330331, India, 
$^{33}$Institute of Environmental Physics, University of Bremen, Bremen, Germany, 
$^{34}$Department of Computer Science, University of Illinois at Urbana-Champaign, 201 N. Goodwin Ave, Urbana, IL 616801, USA, 
$^{35}$Laboratoire Photonique, Num\'erique et Nanosciences UMR 5298, Universit\'e de Bordeaux, Institut d'Optique Graduate School, CNRS, 33400 Talence, France, 
$^{36}$The University of Washington eScience Institute, The University of Washington, Seattle, WA 98105, USA, 
$^{37}$Federal Reserve Board of Governors,  20th and Constitution Ave. NW, Washington, DC 20551, U.S.A., 
$^{38}$Datadog Inc., 620 8th Avenue New York, NY 10018, USA, 
$^{39}$Department of Clinical Sciences Lund, Lund University, Lund, Sweden, 
$^{40}$HQ, Orexplore, Torshamnsgatan 30B, Stockholm, Sweden, 
$^{41}$Statistics Department, University of California - Davis, Davis, CA 95616, USA, 
$^{42}$Emory University, Atlanta, GA 30322, USA, 
$^{43}$Applied BioMath, 561 Virginia Rd Ste 220, Concord, MA 01742, USA, 
$^{44}$Department of Computer Science, University of Oxford, Wolfson Building, Parks Road, OX1 3QD Oxford, UK, 
$^{45}$NeuroInformatics Laboratory, Bruno Kessler Foundation, Trento 38123, Italy, 
$^{46}$Department of Engineering, University of Cambridge, Trumpington St, Cambridge CB2 1PZ, UK, 
$^{47}$Aix Marseille Universit\'e., CNRS, Centrale Marseille, LMA, Marseille, France, 
$^{48}$Interdisciplinary Center for Scientific Computing (IWR), Heidelberg University, Im Neuenheimer Feld 205 69120 Heidelberg, Germany, 
$^{49}$ABB Corporate Research, ABB AG, Wallstadter Str. 59 68526 Ladenburg, Germany, 
$^{50}$Institut f\"ur Mathematische Optimierung, Technische Universit\"at Carolo-Wilhelmina zu Braunschweig, Universit\"atsplatz 2 38106 Braunschweig, Germany, 
$^{51}$Independent Researcher, Cologne, Germany, 
$^{52}$Lawrence Berkeley National Laboratory, Berkeley, CA 94720, USA, 
$^{53}$Institut f{\"u}r Physik, Universit{\"a}t Augsburg, Universit{\"a}tsstra{\ss}e 1, 86135 Augsburg, Germany, 
$^{54}$Applied Research Laboratories, The University of Texas at Austin, P.O. Box 8029 Austin, TX 78713, USA, 
$^{55}$Department of Radiology, School of Medicine, University of Cincinnati, Cincinnati, OH 45221, USA, 
$^{56}$Department of Radiology, Cincinnati Children's Hospital Medical Center, Cincinnati, OH 45229, USA, 
$^{57}$Ascent Robotics Inc., 1-6-10 Hiroo Shibuya Tokyo, Japan, 
$^{58}$ENSTA Bretagne, 2 rue François Verny, Brest, France, 
$^{59}$Faculty of Physics, Ludwig-Maximilians-Universit\"at, Scheinerstr. 1, 81679 M\"unchen, Germany, 
$^{60}$Excellence Cluster Universe, Boltzmannstr. 2, 85748 Garching b. M\"unchen, Germany, 
$^{61}$Independent Researcher, USA, 
$^{62}$Data Sciences, Chan Zuckerberg Biohub, San Francisco, CA 94158, USA, 
$^{63}$Faculty of Mechanical Engineering, University of Ljubljana, A\v{s}ker\v{c}eva 6 1000 Ljubljana, Slovenia-EU, 
$^{64}$Sydney Informatics Hub, The University of Sydney, NSW 2006, Australia, 
$^{65}$Instituto de Astrof\'{i}sica, Pontificia Universidad Cat\'{o}lica de Chile, Casilla 306, Santiago 22, Chile, 
$^{66}$Max Planck Institute for Extraterrestrial Physics, Giessenbachstrasse, 85741 Garching, Germany, 
$^{67}$University of Stuttgart, Machine Learning and Robotics Lab, Universitätsstr. 38, 70569 Stuttgart, Germany, 
$^{68}$Department of Electrical Engineering, Universidade Federal de Campina Grande, 58429900 PB, Brazil, 
$^{69}$Department of Computer Science, Kiel University, 24098 Kiel, Germany, 
$^{70}$Planetary Sciences Group and Florida Space Institute and Department of Physics, University of Central Florida, 4111 Libra Drive Orlando Florida 32816-2385, USA, 
$^{71}$Independent Researcher, Spain, 
$^{72}$Monash Micro Imaging, Monash University, Clayton VIC 3800, Australia, 
$^{73}$Department of Molecular, Cellular, and Developmental Biology, University of Colorado, Boulder, Boulder, CO 80303, USA, 
$^{74}$Department of Physics and Astronomy, Johns Hopkins University, 3400 N. Charles Street, Baltimore, MD 21218, USA, 
$^{75}$Independent Researcher, Munich, Germany, 
$^{76}$University of T\"ubingen, T\"ubingen, Germany, 
$^{77}$Independent Researcher, United Kingdom, 
$^{78}$Two Sigma Investments, 100 Avenue of the Americas 16th Floor, New York, NY 10013, USA, 
$^{79}$Department of Mechanical Engineering, Technical University of Munich, Boltzmannstr. 15, 85748 Garching, Germany, 
$^{80}$Independent Researcher, Berlin, Germany, 
$^{81}$Intel Corp., Austin, TX 78746, USA, 
$^{82}$Independent Researcher, Horten, Norway, 
$^{83}$D. E. Shaw Research, New York, NY 10036, USA, 
$^{84}$Division of Computational Systems Biology, University of Vienna, Althanstra{\ss}e 14, 1090 Vienna, Austria, 
$^{85}$Independent Researcher, Melbourne, Australia, 
$^{86}$School of Physics and Astronomy, University of Manchester, Oxford Road, M13 9PL, UK, 
$^{87}$Electrical and Computer Engineering, University of Wisconsin--Madison, 1415 Engineering Drive Madison, WI, USA, 
$^{88}$Dipartimento di Informatica, Universit\`a degli Studi di Milano, via Celoria 18, 20133 Milano MI, Italy, 
$^{89}$Inter-University Centre for Astronomy and Astrophysics, Ganeshkhind, Pune, 411007, India, 
$^{90}$Kavli Institute for the Physics and Mathematics of the Universe, Kashiwanoha 5-1-5, Kashiwa-shi, 2778583, Japan, 
$^{91}$Waymo LLC, Mountain View, CA 94043, USA, 
$^{92}$Faculty of Engineering, Niigata University, 8050 Ikarashi-Ninocho, Nishi-ku, Niigata, 9502181, Japan, 
$^{93}$Virginia Polytechnic Institute and State University, Blacksburg, VA 24061, USA, 
$^{94}$Aperio Software, Headingley Enterprise and Arts Centre, Bennett Road, Leeds, LS6 3HN, United Kingdom, 
$^{95}$Broad Institute, Cambridge, MA 02142, USA, 
$^{96}$Department of Psychology, Humboldt University of Berlin, Rudower Chaussee 18, 12489 Berlin, Germany, 
$^{97}$Epiq Solutions, Schaumburg, IL 60173, USA, 
$^{98}$Max Planck Institute for Software Systems, Saarbr\"ucken, Germany, 
$^{99}$Department of Psychology and Brain Sciences, Dartmouth College, Hanover, NH 03755, USA, 
$^{100}$Jacobs School of Engineering, University of California San Diego, 9500 Gilman Drive, La Jolla, CA 92161, USA